\documentclass[prd,aps,preprint]{revtex4}
\usepackage{graphicx}
\begin{document}
\tolerance=5000
\def\be{\begin{equation}}
\def\ee{\end{equation}}
\def\bea{\begin{eqnarray}}
\def\eea{\end{eqnarray}}
\def\ii{\'{\i}}
\def\bi{\bigskip}
\def\be{\begin{equation}}
\def\en{\end{equation}}
\def\bq{\begin{eqnarray}}
\def\eq{\end{eqnarray}}
\def\noi{\noindent}
\title{Pioneer's Anomaly and the Solar Quadrupole Moment}
\author{Hernando Quevedo}
\email{quevedo@nuclecu.unam.mx}
\affiliation{Instituto de Ciencias Nucleares\\
Universidad Nacional Aut\'onoma de M\'exico\\
A.P. 70-543,  M\'exico D.F. 04510, M\'exico\\}

\date{\today}
\begin{abstract}
\setlength{\baselineskip}{.5cm}
The trajectories of test particles moving in the gravitational field of a non-spherically
symmetric mass distribution become affected by the presence of multipole moments.
In the case of hyperbolic trajectories, the quadrupole moment of an oblate mass induces a
 displacement of the trajectory towards the mass source, an effect that can be 
 interpreted as an additional acceleration directed towards the source. Although this additional
 acceleration is not constant, we perform a general relativistic analysis in order to
 evaluate the possibility of explaining Pioneer's anomalous
  acceleration by means of the observed
Solar quadrupole moment, within the range of accuracy of the observed anomalous acceleration.
We conclude that the Solar quadrupole moment generates an acceleration which is of the same
order of magnitude of Pioneer's constant acceleration only at distances of a few astronomical
units.

\end{abstract}
\maketitle
%\setlength{\baselineskip}{1\baselineskip}

%\newpage

\section{INTRODUCTION}
\label{sec:int}

Pioneer 10/11 are spacecraft which were
launched more than thirty years ago to explore the outer region of the Solar 
system. 
They were also the first missions to enter the edge of interstellar space. 
Due to their especial structure, they also represent an ideal system to 
carry out precision celestial mechanics experiments which permit acceleration
estimations to the level of $10^{-8}cm/s^2$. After encountering the outer planets
Jupiter and Saturn,
the two spacecraft followed hyperbolic orbits near the plane of the ecliptic
to opposite sides of the Solar system. 
A careful analysis of orbital 
data from Pioneer 10/11 has been reported in \cite{obs1,obs2} which indicates 
the existence of a very weak, long-range acceleration, 
$a_P = (8.74\pm 1.33)\times 10^{-8} cm/s^2$, directed towards the Sun. 
The most intriguing result about this acceleration is that it is constant, with 
a single measurement accuracy averaged over 5 days. Moreover, it has been measured
at distances contained within the approximate 
range of 20 to 60 astronomical units (AU) from the Sun.

The most conservative
hypothesis about the origin 
of this anomalous acceleration is that this kind of effects 
on Pioneer's tracking data must be due to some technological reasons
related to the configuration and equipment of the spacecraft. However,
the analysis reported in \cite{obs1,obs2} seems to have ruled out
all possible non-gravitational forces generated inside and outside
the spacecraft. Moreover, the authors have taken into account the
accepted values of the errors in the determination of the planetary 
ephemeris, Earth's orientation, precession, and nutation. 
More adventurous explanations of the anomalous acceleration 
suggest the existence of new physical phenomena, especially 
inspired by the fact that $a_P$ is of the order of $cH$, where
$c$ is the velocity of light in empty space and $H$ is the 
Hubble constant. For instance, in \cite{san,guru,ost} it is 
suggested 
that the expansion of the Universe induces local tidal forces
which generate the anomalous acceleration. In \cite{bel} 
a Kaluza-Klein model is proposed in which the fifth dimension 
becomes dynamical with an expansion rate that generates $a_P$. 
A scale dependent cosmological term has been used in \cite{mod}
which affects the motion of test particles. A different explanation
is suggested in \cite{man} by means of a time-dependent Newtonian
gravitational constant that produces a long-range acceleration. 
More sophisticated explanations have been proposed in the 
literature (see, for instance \cite{obs2}, for a recent review)
which include dark matter and stringy inspired scalar fields \cite{bert1,bert2}.

In this work, we investigate the possibility of explaining
the anomalous acceleration by means of the Solar quadrupole
moment. Clearly, the additional acceleration due to the 
quadrupole is not constant, but decreases with distance. 
In this sense the, quadrupole moment could immediately be ruled out
as an explanation for $a_P$. However, our approach is different.
Since $a_P$ has been observed only at distances less than 60 AU,
with an accuracy of around 15\%, our goal is to determine 
whether  the acceleration generated by the quadrupole is of the
same order of $a_P$ and can be fitted within its 
range of accuracy. This idea is based upon the fact that the
quadrupole moment of the Sun has been determined in 
an exact manner only recently \cite{roz}, and has been given
the value of $J_2 = - 2.28 \ 10^{-7}\ \pm 15$\%, where $J_2$ 
is a dimensionless parameter to be introduced below. 
Before this value was known, different methods generated 
completely different values of $J_2$, sometimes with 
differences of several orders of magnitude. (In fact, 
such a great ambiguity in the value of $J_2$ could easily 
be used to explain $a_P$ in the complete range of observation!)
With this new value, we will show that the quadrupole moment
of the Sun
produces an acceleration which is of the same order of $a_P$ 
at distances of a few AU's, but is several orders of
magnitude less than $a_P$ at distances between 20 and 60 AU's.

We will analyze the relativistic hyperbolic motion of test particles in the
gravitational field of a mass with quadrupole moment. For the
sake of simplicity, we will assume that the quadrupole is due
only to the asphericities of the mass distribution, preserving
the axial symmetry, and will neglect the rotation of the source.
This allows us to use a static, axisymmetric solution of Einstein's
vacuum field equations as the metric which describes the spacetime.
In particular, we will use the Erez--Rosen metric for our analysis.
Moreover, we will assume that the motion is confined to the 
equatorial plane of the source. This is in good agreement with 
the actual path of Pioneer 10/11 which move near the plane of the 
ecliptic.

\section{Gravitational field of a mass with quadrupole moment}
\label{sec:erro}

In this section we present a static axisymmetric solution of
Einstein's vacuum field equations which was first discovered
by Erez and Rosen \cite{erro}. This solution generalizes
Schwarzschild solution to include an arbitrary mass quadrupole
moment. Originally, this solution was obtained in prolate
spheroidal coordinates, but for the purposes of this work
spherical coordinates $(t, r, \theta, \varphi)$
are more appropriate. In these coordinates, it can be shown that
it is symmetric with respect to the equatorial plane,
$\theta=\pi/2$, so that it becomes a geodesic plane. For the
sake of simplicity we will limit ourselves to the investigation
of free motion on the equatorial plane. Accordingly, the
Erez-Rosen solution can be written as \cite{fort}
\be
ds^2 = \left( 1 - \frac{2m}{r}\right) e^{2q\psi} dt^2
- \left( 1 - \frac{2m}{r}\right)^{-1} e^{2q(\gamma - \psi)} dr^2
 - r^2 e^{-2q\psi} d\varphi^2 \ ,
\label{erromet}
\ee
where the constant $m$ represents the Schwarzschild mass,
$q$ is a constant parameter that determines the quadrupole moment,
and $\psi$ and $\gamma$ are, in general, functions of $r$ and $\theta$.
On the equatorial plane, these functions take the form
\be
\psi = - \frac{1}{4}\left[\frac{1}{2}\left( 3\frac{r^2}{m^2} - 6 \frac{r}{m}
+2\right)\ln\left(1 -\frac{2m}{r}\right) + 3\left(\frac{r}{m} -1\right)\right] \ ,
\ee
\be
\gamma \approx \frac{1}{10}\frac{m^4}{r^4} \ .
\ee
The general form of the function $\gamma$ is rather cumbersome. Here we
quote only the leading term in the limit $r\rightarrow\infty$ which is sufficient
for the analysis we will perform in the next section.

In general, the Erez-Rosen
solution is asymptotically flat and this allows us to determine its multipole
moments in a covariant manner. For instance, using the Geroch-Hansen \cite{gerhan}
definition of multipole
moments one can show that the monopole  is $m$ and the quadrupole moment is
$(2/15)qm^3$. Higher multipoles are given in terms of the monopole and quadrupole moments.
So the parameters $m$ and $q$ acquire a clear physical meaning.
As expected, in
the limiting case $q=0$, the Erez-Rosen metric (\ref{erromet}) coincides with the
Schwarzschild solution on the equatorial plane. The hypersurface $r=2m$ turns out
to be singular, in accordance with the  black hole uniqueness theorems. An additional
singularity is situated at $r\rightarrow 0$. Outside the hypersurface $r=2m$, the
spacetime is completely regular.

At large distances from the source, the expansion of the metric component $g_{tt}$
leads to 
\be
g_{tt}\approx
 1 - \frac{2m}{r}\left[ 1 - \frac{2}{15} q \left(\frac{m}{r}\right)^2\right] \ .
\label{gtt}
\ee
Later on we will use this expression to determine the acceleration generated by the 
quadrupole moment.

\section{Motion of test particles}
\label{sec:motion}

Let us consider the motion of test particles on the equatorial plane of the
Erez-Rosen metric (\ref{erromet}). The geodesic equation can be reduced to a
set of first order differential equations due to the existence of two constants
of motion $p_t$ and $p_\varphi$ associated with the time-translation
and the axial symmetry, respectively. For the case of a massive test particle,
the geodesic equations can be written as:
\be
\left(1-\frac{2m}{r}\right) e^{2q\psi} \dot t = p_t \ ,
\ee
\be
r^2e^{-2q\psi}\dot \varphi = p_\varphi \ ,
\ee
\be
\dot r ^2 = e^{-2q\gamma} \left[ p_t^2 - \left(1-\frac{2m}{r}\right) e^{2q\psi}
- \left(1-\frac{2m}{r}\right)\frac{p_\varphi^2}{r^2}e^{4q\psi}\right] \ ,
\label{geor}
\ee
where a dot represents the derivative with respect to the affine parameter
along the geodesic. It is possible to perform a detailed analysis of these
set of geodesic equations \cite{fort} and find out all the effects of the
quadrupole moment on the motion of test particles. Here, however, we are
interested only on hyperbolic motion. Hence, let us consider the
special case $r=r(\varphi)$ so that $\dot r= (dr/d\varphi) \dot \varphi$,
and introduce the following notations:
\be
u = \frac{2m}{u}\ , \quad \alpha = \frac{2m}{p_\varphi} \ .
\label{defu}
\ee
Then, the geodesic equation (\ref{geor}) reduces to
\be
(u')^2 = F(u):=e^{-2q\gamma}\left[ u^3 - u^2 + \alpha^2 u e^{-2q\psi} +
\alpha^2 e^{-2q\psi} \left( p_t^2 e^{-2q\psi} -1\right) \right] \ ,
\label{geou}
\ee
where a prime denotes the derivative with respect to the angle coordinate
$\varphi$. Since the expression $(u')^2$ is positive for a timelike geodesic,
we must demand that the condition $F(u)\geq 0$ be satisfied.
Accordingly, from Eq.(\ref{geou}) we obtain
\be
\alpha^2 p_t^2 \geq (1-u) e^{2q\psi}\left( \alpha^2 + u^2 e^{2q\psi}\right) \ .
\ee
Because the left-hand side of this inequality is a positive definite constant,
this implies that the roots of the equation $F(u)=0$ must be contained
in the interval $(1-u)>0$ or, according to Eq.(\ref{defu}), in the
interval $r\in (2m,\infty)$. As mentioned in the last section, the
hypersurface $r=2m$ contains in general a curvature singularity. Thus, hyperbolic
motion is confined to the region where the Erez-Rosen metric is regular.

On the other hand, simple algebraic manipulations show that the
equation $F(u)=0$ can be expressed as a quadratic algebraic equation
for the function $\exp(2q\psi)$, i.e.,
\be
e^{4q\psi} +\frac{\alpha^2}{u^2}  e^{2q\psi} - \frac{\alpha^2 p_t^2}{(1-u)u^2} =0 \ .
\ee
Since $\exp(2q\psi)>0$, we need to consider only the positive
root of this equation which can be expressed as
\be
q = \frac{1}{2\psi}\ln \left [\frac{\alpha^2}{2u^2}\left(
\sqrt{1 + 4 \frac{p_t^2}{\alpha^2}\frac{u^2}{1-u} } -1 \right)\right] \ .
\label{qeq}
\ee
Now, the function $\psi$ is positive in the interval $r\in (2m,\infty)$.
Then, from Eq.(\ref{qeq}) we obtain that
\be
\frac{\alpha^2}{2u^2}\left(
\sqrt{1 + 4 \frac{p_t^2}{\alpha^2}\frac{u^2}{1-u} } -1 \right) > 1 \quad
{\rm for} \quad q> 0 \ ,
\ee
\be
\frac{\alpha^2}{2u^2}\left(
\sqrt{1 + 4 \frac{p_t^2}{\alpha^2}\frac{u^2}{1-u} } -1 \right) < 1 \quad
{\rm for} \quad q< 0 \ .
\ee
These conditions can be represented in an equivalent manner as
\be
f(u):= u^2(u-1) + \alpha^2 (u-1) + \alpha^2 p_t^2  > 0 \quad {\rm or}
\quad < 0 \quad {\rm for} \quad q>0 \quad {\rm or }\quad q<0 \ ,
\label{conds}
\ee
respectively.
On the other hand, as can be seen from Eq.(\ref{geou}), the function $f(u)$
coincides with the function $F(u)$ for vanishing quadrupole moment $(q=0)$,
i.e., $f(u)$ determines the motion of test particles
in the Schwarzschild spacetime. This is a nontrivial result. In fact,
suppose that the equation $F(u) =0$ has a root at  $u= u_0$. Then at $u_0$
the value of $q$ is related to the value of $u_0$ by means of (\ref{qeq})
with $u$ replaced by $u_0$. If, for instance, $q<0$, then according
to Eq.(\ref{conds}) the Schwarzschild function $f(u_0)$ is negative.
In the case of hyperbolic motion, the value $u=u_0$ corresponds to a
vanishing ``velocity", i.e., $u'(u=u_0) =0$ or, equivalently, $\dot r =0$,
a condition which can be satisfied only at the perihelion of the hyperbolic
trajectory. At this point, the motion of test particles in the
Schwarzschild spacetime is not allowed, because there $f(u_0)<0$.
From the form of the functions $F(u)$ and $f(u)$ one can show that
in general
\be
f(u)<F(u) \quad {\rm for} \quad q<0 \qquad {\rm and} \quad
f(u)>F(u) \quad {\rm for} \quad q>0 \ .
\label{beh}
\ee

Let us consider the case of an oblate configuration $(q<0)$. Hyperbolic
motion corresponds to the case where $F(u=0)>0$ and possesses (at least) one
positive root, say $u_2$.
 The motion is confined to the region $0<u<u_2$
with $r_2 = 2m/u_2$ being the perihelion distance.
At $u=u_2$, according to (\ref{beh}), we have that $f(u_2)<0$ and so
no motion is allowed in the Schwarzschild spacetime. If we ``move" from
$u_2$ towards $u=0$, we will find a point, say $u_1$, where $f(u_1)=0$.
Starting from this point and up to the point $u=0$, where $ f(u=0)>0$,
the motion is allowed in the Schwarzschild as well as in the Erez--Rosen
spacetime. For the trajectory in the Schwarzschild spacetime the perihelion
distance is given by $r_1 =2m/u_1$ and at this point, according to
(\ref{beh}), $F(u=u_1) > 0$, that is, the trajectory in the Erez--Rosen
spacetime would correspond to a non-zero radial velocity. So we see that
the perihelion distance of a hyperbolic path becomes affected by the
presence of an oblate quadrupole moment in such a way that
\be
r_2 < r_1 \ ,
\ee
a result that trivially follows from the fact that $u_1<u_2$. Thus an oblate
quadrupole moment reduces the perihelion distance.

Let us denote by $r_{ER}(\varphi)$
the hyperbolic path of a test particle in the Erez--Rosen spacetime with
mass $m$ and a negative quadrupole $q$, and by $r_S(\varphi)$
the corresponding path
in the Schwarzschild spacetime which is obtained from $r_{ER}(\varphi)$ by
setting $q=0$. Notice that the mass $m$ must be the same in order to be able
to compare both trajectories. Clearly, $r_S(\varphi)$ must be a solution
of the equation $(u')^2 = f(u)$, whereas $r_{ER}(\varphi)$ is a solution of
$(u')^2 = F(u)$ with the same mass. As we have shown above, at the perihelion
we have that $r_{ER}(\varphi_{per}) < r_S(\varphi_{per})$. Moreover, since
in this case $f(u)<F(u)$ for all values of $u$, we obtain that in general
\be
r_{ER}(\varphi) < r_S(\varphi)
\ee
at all points of the trajectory. The difference between these two trajectories
will become smaller as $r$ increases. Moreover, $r_{ER}$ will approach $r_S$
asymptotically at $r\rightarrow\infty$, where the contribution of the
mass quadrupole moment becomes negligible. If we were analyzing the motion
of a test particle in the gravitational field of a mass with negative
quadrupole moment,  ignoring the contribution of the quadrupole, we would
expect a path $r_S(\varphi)$, whereas the measured path would be
$r_{ER}(\varphi)$, which is always less than $r_S(\varphi)$.
Then, one could interpret this result as due to an additional
acceleration directed towards the source of the gravitational field.
In fact, the presence of a quadrupole moment generates an ``effective"
acceleration which leads to deformations of the trajectories of 
test particles. In the next section we will estimate this effective
acceleration.

To finish this section we mention that in the case of
a prolate quadrupole moment $(q>0)$ the effect is  opposite, 
that is, the perihelion distance increases and the effective
acceleration would be directed outwards the gravitational source. 
This case, however, is not expected to occur in astrophysical bodies
because any deviations from spherical symmetry are usually generated
by rotation which leads to oblate configurations.

\section{The effective acceleration}
\label{sec:acc}

To find the exact hyperbolic orbit of a test particle in the 
gravitational field described by the Erez--Rosen metric one must
solve the differential equation (\ref{geou}) in the range 
where this kind of motion is allowed. This can be done by applying
standard numerical methods. Another possibility is to consider
the case of motion at large distances from the source, where we can
use a Taylor
expansion of the functions $\psi$ and $\gamma$ that enter the function
$F(u)$ in Eq.(\ref{geou}), and preserve the leading terms only.
The resulting function $F(u)$ turns out to be a cubic polynomial 
of $u$. In this case, the differential equation (\ref{geou}) can
be solved analytically as $\varphi=\varphi(u)$ (see \cite{linda} for 
an analysis of the resulting solutions), an expression from which, 
in principle, we could derive the acceleration due to the quadrupole
moment $q$. Nevertheless, the resulting expression is rather cumbersome
and an analysis seems to be possible only by applying numerical procedures.

To evaluate the acceleration $a_q$ induced by the quadrupole moment, we 
will use the analogy with the method applied
 to measure the Solar quadrupole. Indeed, for an axially symmetric
distribution of matter, for instance the Sun, 
the contribution of the quadrupole
moment to the exterior gravitational field can be expressed in 
the metric component $g_{tt}$ as
\be
g_{tt} = 1 - 2\frac{m}{r}\left[ 1 + J_2 \left(\frac{R_0}{r}\right)^2
 P_2(\cos\theta) \right] \ ,
\label{gtt1}
\ee
where $J_2$ is a dimensionless constant, $P_2(\cos\theta)$ is 
the Legendre polynomial, and $R_0$ is the radius of the Sun. 
Evaluating this expression on the equatorial
plane $(\theta=\pi/2)$ and comparing it with Eq.(\ref{gtt}), 
we can see that both expressions coincide if 
\be
q = 60 J_2 \left(\frac{R_0}{r}\right)^2 \ .
\ee
Then, from Eq.(\ref{gtt1}) we can read the expression for the 
gravitational potential whose derivative with respect to the 
radial coordinate produces the leading term of the acceleration.
Thus, we obtain
\be
a_q = -\frac{3J_2 R_0 m}{2r^4} \ .
\label{aq}
\ee
For the evaluation of this quantity we take, according to \cite{roz}, 
the value of  
$J_2 = - 2.28 \ 10^{-7}$ for the Solar quadrupole 
moment and $R_0 \approx 6.96 \ 10^{12} cm$ for the photospheric radius of
the Sun. Moreover, we express the distance $r$ in astronomical units, 
i.e., $r = n \ 1.496 \ 10^{13}cm$, where $n$ is a positive real number. 
Thus, we obtain
\be
a_q = \frac{1}{n^4}\  4.33\ 10^{-8} cm/s^2 \ .
\ee
So we see that at a distance of one astronomical unit, the order 
of magnitude of $a_q$ coincides with $a_P$. Considering the uncertainty 
in the values of $J_2$, $R_0$ and $m$, one can show that $a_q$ and $a_P$
are still of the same order of magnitude at distances smaller than 
$n\approx 2.1$.  In the interval between 20 and 60 AU's, 
where the anomalous acceleration
has been measured, $a_q$ is several orders of magnitude less than $a_P$.

\section{CONCLUSIONS}
\label{sec:conclusions}

In this work we have investigated the possibility of explaining
Pioneer's anomalous acceleration by means of the Solar quadrupole
moment. In the analysis of the hyperbolic motion of a test particle
in the gravitational field of a mass with a negative quadrupole moment, 
we have
shown that the complete trajectory becomes closer to the gravitational
source, an effect that can be interpreted as due to the presence of an
additional acceleration, $a_q$,  generated by the quadrupole moment.

Although $a_q$ is not a constant quantity, its value at distances very 
closed to one AU is of the same order of magnitude of $a_P$. 
The discrepancy between $a_q$ and $a_P$ increases drastically with distance,
indicating that it is not possible to explain the anomalous
acceleration at distances where it has been measured for Pioneer 10/11
spacecraft.

\section*{ACKNOWLEDGMENTS}

This work was supported by DGAPA-UNAM grant IN112401 and CONACyT-Mexico grant 36581-E.


\begin{thebibliography}{99}
%\begin{references}


\bibitem{obs1} J. D. Anderson et al., Phys. Rev. Lett. {\bf 81}, 2858 (1998);
S. G. Turyshev et al., gr-qc/9903024.

\bibitem{obs2} J. D. Anderson et al., Phys. Rev. D
 {\bf 65}, 082004 (2002).

\bibitem{san} J. L. Rosales and J. L. S\'anchez-Gomez, gr-qc/9810085.

\bibitem{guru} V. Guruprasad, astro-ph/9907363, gr-qc/0005014, gr-qc/0005090.

\bibitem{ost} D. Ostvang, gr-qc/9910054.

\bibitem{bel} W. B. Belayev, gr-qc/9903016.

\bibitem{mod} G. Modanese, Nucl. Phys. B {\bf 556}, 397 (1999).

\bibitem{man} R. Mansouri, F. Nasseri, and M. Khorammi, Phys. Lett. A
{\bf 259}, 194 (1999).

\bibitem{bert1} O. Bertolami and J. Paramos, Class. Quantum Grav. {\bf 21}, 
3309     (2004). 
\bibitem{bert2} O. Bertolami and J. Paramos, gr-qc/0411020.


\bibitem{roz} J. P. Rozelot, S. Pireaux, S. Lefebvre, and T. Corbard,
astro-ph/0403382.



\bibitem{erro}
G. Erez and N. Rosen, Bull. Res. Council of Israel
{\bf 8F}, 47 (1959);
Ya. B. Zeldovich and I. D. Novikov, {\it Relativistic Astrophysics} (University
of Chicago Press, Chicago, 1971).

\bibitem{fort}
H. Quevedo, Fortschr. Phys. {\bf 38}, 733 (1990).

\bibitem{gerhan} R. Geroch, J. Math. Phys. {\bf 11}, 2580 (1970); R. O. Hansen, J. Math.
Phys. {\bf 15}, 46 (1974).

\bibitem{linda} H. Quevedo and L. Parkes, Gen. Rel. Grav. {\bf 21}, 1047 (1989).





\end{thebibliography}
\end{document}